\documentclass{ws-procs9x6}

\def\simgt{\,\rlap{\lower 3.5 pt\hbox{$\mathchar \sim$}}\raise 1pt \hbox {$>$}\,}
\def\simlt{\,\rlap{\lower 3.5 pt\hbox{$\mathchar \sim$}}\raise 1pt \hbox {$<$}\,}

\def\psibarpsi{\overline{\psi}\psi}

\newcommand{\dslash}{D \!\!\!\! /}

\begin{document}

\title{QCD at non-zero chemical potential and temperature from the lattice}

\author{C.R. Allton\footnote{Speaker at the Workshop}}

\address{Department of Physics, University of Wales Swansea,
 Singleton Park, Swansea, SA2 8PP, U.K.\footnote{Permanent Address}\\
{\rm and} \\
Department of Mathematics, University of Queensland,
Brisbane 4072, Australia}

\author{S. Ejiri, S.J. Hands}
\address{Department of Physics, University of Wales Swansea,
          Singleton Park, Swansea, SA2 8PP, U.K.}

\author{O. Kaczmarek, F. Karsch, E. Laermann, Ch. Schmidt}

\address{Fakult\"at f\"ur Physik, Universit\"at Bielefeld,
          D-33615 Bielefeld, Germany}

\author{L. Scorzato}

\address{Department of Physics, University of Wales Swansea,
          Singleton Park, Swansea, SA2 8PP, U.K. \\
{\rm and} \\
DESY Theory Division, Notkestrasse 85,
D-22603 Hamburg, Germany.}

\maketitle

\abstracts{
A study of QCD at non-zero chemical potential, $\mu$, and temperature,
$T$, is performed using the lattice technique. The transition
temperature (between the confined and deconfined phases) is determined
as a function of $\mu$ and is found to be in agreement with other
work. In addition the variation of the pressure and energy density
with $\mu$ is obtained for $\mu \simgt 0$. These results are of particular
relevance for heavy-ion collision experiments.
}

%}}}

%{{{ introduction

\section{Introduction}

The QCD phase diagram has come under increasing experimental and
theoretical scrutiny over the last few years. On the experimental
side, very recent studies of compact astronomical objects have
suggested that their cores contain ``quark matter'', i.e. QCD in a
new, unconfined phase where the basic units of matter are quarks,
rather than nuclei or nucleons\cite{quarkstars}. More terrestrially,
heavy ion collision experiments, such as those performed at RHIC and
CERN, are also believed to be probing unconfined QCD\cite{hic}. On the
theoretical side, the study of QCD under these extreme densities and
temperatures has proceeded along several fronts.  One of the most
promising areas of research is the use of lattice techniques to
study either QCD itself, or model theories which mimic the strong
interaction\cite{simon}. Clearly the most satisfying approach would be
the former, i.e. a direct lattice study of QCD at various coordinates
$(T,\mu)$ in its phase space ($\mu$ is the chemical potential for
the quark number).
However, until recently, this has proved intractable at a practical
level for very fundamental reasons. This is because the Monte Carlo
integration technique, which is at the heart of the (Euclidean)
lattice approach, breaks down when $\mu \ne 0$.  This work summarises
one new approach which overcomes this problem and has made progress
for $\mu \ne 0$ and $T \ne 0$.

In the next section a summary is given of the lattice technique
and the problem incurred when $\mu\ne 0$.
Section \ref{Reweighting} describes the method used to overcome
these difficulties and section \ref{Simulation} outlines the
simulation details. The next two sections apply the method to
variations in $m$ and $\mu$, and section \ref{Pressure} describes
calculations of the pressure and energy density as functions
of $\mu$.

A full account of this work is published elsewhere\cite{us}.

%}}}

%{{{ Lattice technique

\section{Lattice technique}

On the lattice, the quark fields, $\psi(x)$, are defined on the sites, $x$,
and the gluonic fields, $U_\mu(x)$, on the links $x\rightarrow x+\mu$.
Observables are then calculated via a Monte Carlo integration approach:
\begin{equation} \label{omega}
<{\Omega}> \;\;=\;\; \frac{1}{N_{\{U,\psi,\bar{\psi}\}}}
\sum_{\{U,\psi,\bar{\psi}\}}\ ^{\!\!\!\!\!\!\!\!'}
\;\;{\Omega}(\psi ,\bar{\psi} ,U)
\end{equation}
where $\sum '$ represents a sum over configurations
$\{U,\psi,\bar{\psi}\}$ which are selected with probability
proportional to the Boltzmann weight $P(\{U,\psi,\bar{\psi}\}) \propto
e^{-S(\{U,\psi,\bar{\psi}\})}$ where $S$ is a suitably defined (Euclidean)
gauge-invariant action.
The fermionic part of this action is
\begin{eqnarray}
S_F = \sum_x \bar{\psi}(x) &\underbrace{(\dslash + m)}& \psi(x).
 \\ \nonumber
 &{=M}&
\end{eqnarray}
For $\mu=0$ it can be shown that this action produces a (real-valued) positive
Boltzmann weight.

Calculations at non-zero temperature, $T\ne 0$,
can be performed by using a lattice with a finite temporal extent
of $N_t a = 1/T$, where $N_t$ is the number of
lattice sites in the time dimension.
In practice, $T$ is varied by changing the gauge coupling,
$g_0$, and hence (through dimensional transmutation) the lattice
spacing, $a$, rather than by changing $N_t$ (which can only be
changed in discrete steps!).

The chemical potential is introduced into the system via an additional
term in the quark matrix $M$, proportional to the Dirac matrix, $\gamma_0$,
\begin{equation}
{M} \rightarrow {M} + \mu \gamma_0.
\end{equation}
For $\mu \ne 0$, this leads to a complex-valued Boltzmann ``weight''
which can therefore no longer be used as a probability distribution,
and, hence, the Monte Carlo integration procedure is no longer applicable.
This is known as the {\em Sign Problem} and has plagued more than
a decade of lattice calculations of QCD at $\mu \ne 0$.

%}}}

%{{{ Reweighting

\section{Reweighting}
\label{Reweighting}

This section outlines the Ferrenberg-Swendsen {\em reweighting}
approach\cite{fs} which is used to overcome the sign problem detailed
in the previous section. Observables at one set of parameter values
$(\beta,m,\mu)$ (where $\beta=6/g_0^2$, and $m$ is the quark mass) can
be calculated using an ensemble generated at another set of parameters
$(\beta_0,m_0,\mu_0)$ as follows,

\begin{eqnarray}
\langle {\Omega} \rangle_{(\beta, m, \mu)}
&\equiv& \frac{1}{{Z}(\beta, m, \mu)} \int
{D}U {\Omega} \det M(m, \mu) {\rm e}^{-S_g(\beta)}
\label{eq:rew} \\
&=& \frac{\left\langle {\Omega} \;{\rm e}^{
(\ln \det M(m, \mu)-\ln \det M(m_0,\mu_0))} {\rm e}^{-S_g(\beta)+S_g(\beta_0)}
 \right\rangle_{(\beta_0, m_0, \mu_0)} }{
\left\langle{\rm e}^{
(\ln \det M(m, \mu)-\ln \det M(m_0,\mu_0))} {\rm e}^{-S_g(\beta)+S_g(\beta_0)}
  \right\rangle_{(\beta_0, m_0, \mu_0)}}.
\label{eq:opexp}
\end{eqnarray}
Here $S_g$ is the gauge action.

In principle, Eq.(\ref{eq:opexp}) can be used to map out the entire
phase diagram of QCD. However, it has been found that its naive
application fails since the relative size of fluctuations in both
numerator and denominator tend to grow exponentially with the volume
of the system studied. This is a signal that the overlap between the
ensemble at $(\beta_0, m_0, \mu_0)$ is small compared with that at
$(\beta, m, \mu)$.

One study which has had success in using Eq.(\ref{eq:opexp}) is that
of Fodor and Katz\cite{fodorkatz}. They apply the reweighting approach
not at an arbitrary point, $(T,\mu)$, in the phase diagram, but rather
they trace out the phase transition line $T_c(\mu)$.  This method is
successful presumably because the overlap between the ensembles
remains high along the ``coexistence'' line which defines the
transition.

This paper utilises an alternative approach which Taylor expands
Eq.(\ref{eq:opexp}) as a function of $\mu$ (or $m$) and hence
estimates the derivatives of various quantities w.r.t. $\mu$ (or
$m$). Derivatives up to the second order are considered, thus for
$<\Omega>$ (in the case of a Taylor expansion in $\mu$) we have
\begin{equation}
\langle{\Omega}\rangle_{(\beta,\mu)}=
{\langle({\Omega}_0+
 {\Omega}_1\mu+{\Omega}_2\mu^2)
  \exp({R}_1\mu+{R}_2\mu^2-\Delta S_g)\rangle_{(\beta_0,\mu_0)}
\over
{\langle\exp({R}_1\mu+{R}_2\mu^2-\Delta S_g)
\rangle_{(\beta_0,\mu_0)}}},
\label{eq:specific}
\end{equation}
where $\mu_0 = 0$.
In Eq.(\ref{eq:specific}), ${R}_n$ is the $n-$th derivative of the
fermionic reweighting factor in Eq.(\ref{eq:opexp}), i.e.
\begin{eqnarray}
\label{eq:expand}
\ln\left({{\det M(\mu)}\over{\det M(0)}}\right) =
\sum_{n=1}^{\infty} \frac{\mu^n}{n!}
\frac{\partial^n \ln \det M(0)}{\partial \mu^n}\equiv\sum_{n=1}^\infty
{R}_n\mu^n.
\end{eqnarray}
The $\Omega_n$ are similarly the $n-$th derivatives of the observable $\Omega$.

Two observables are studied: the chiral condensate $<\psibarpsi>$
and the Polyakov Loop, $L$.
Since $L$ is a pure gluonic quantity, defined as
\begin{equation}
<L> = < \frac{1}{V_s}\sum_{\vec{x}} Tr \prod_{t=0}^{N_t} U_t(\vec{x},t) >,
\end{equation}
all of its derivatives are zero. (Here, $V_s$ is the spatial volume.)
However, the expansion of $<\psibarpsi>$ is more challenging since it
is defined as
\begin{equation}
\langle \bar{\psi} \psi \rangle
\sim  \langle {\rm tr} M^{-1} \rangle,
\label{eq:psibarpsi}
\end{equation}
and hence the application of Eq.(\ref{eq:specific}) requires
determinations of $\partial^n M^{-1} / \partial \mu^n$.

The susceptibilities, $\chi$, of both $<\psibarpsi>$ and $L$ are
defined as usual by their fluctuations, e.g.
\begin{equation}
\chi_L = \mbox{(volume factor)} \; \times \; (<L^2> - <L>^2).
\label{eq:sus}
\end{equation}
These susceptibilities have a maximum at the transition point,
$\beta_c$, and hence can be used to determine the transition
point $\beta_c(m,\mu)$.

%}}}

%{{{ Simulation details

\section{Simulation details}
\label{Simulation}

The lattice calculations were performed using a ``p4-improved''
discretisation of the continuum action which is a sophisticated
lattice action maintaining rotational invariance of the free fermion
propagator up to $O(p^4)$ \cite{Kar00,Hel99} ($p$ is the
momentum here).  Two dynamical flavours of quarks were used with a
$16^3 \times 4$ lattice.  Simulations were performed at quark mass, $m
= 0.1$ and $0.2$ which correspond to (unphysically heavy)
pseudoscalar-vector meson mass ratios of $M_{PS}/M_V \approx 0.70$ and
$0.85$. Approximately 400,000 configurations were generated in total
using around 6 months of a 128-node (64Gflop peak) APEmille in the
University of Wales Swansea.

%}}}

%{{{ Results for mass reweighting

\section{Results for mass reweighting}

As a check of our method, we first use mass reweighting, since, unlike
the $\mu \ne 0$ case, there is no theoretical difficulty in simulating
at virtually any value of $m$, and hence there are published data at a
number of different $m$ values readily available for comparison.
Reweighting in quark mass is simply a matter of setting $\mu=\mu_0=0$
in Eq.(\ref{eq:opexp}) and Taylor expanding in $m$ rather than $\mu$
in Eqs.(\ref{eq:specific} \& \ref{eq:expand}).  (Note that $\partial
M/\partial m = 1$ and $\partial^n M/\partial m^n = 0$ for $n>1$.)

\begin{figure}[th]
\centerline{\epsfxsize=3.2in\epsfbox{
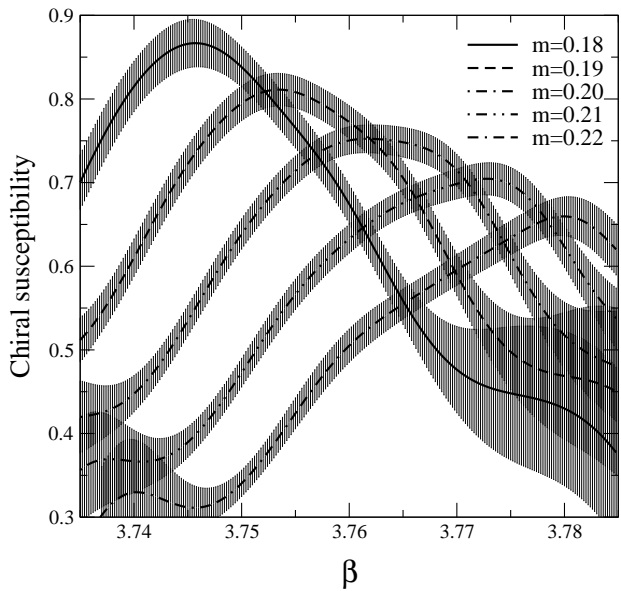
}}
\caption{The quark mass dependency of $\chi_{\psibarpsi}$
as a function of $\beta$
in the neighbourhood of $m=0.2$ for the Chiral Susceptibility.
\label{rew_m}}
\end{figure}

We use the peak position in both the chiral condensate and Polyakov
Loop susceptibilities, $\chi_{\psibarpsi,L}$ to determine the phase
transition point $\beta_c(m)$. Figure \ref{rew_m} illustrates this by
plotting $\chi_{\psibarpsi}$ as a function of $\beta$ (for $m=0.2$).
This shows the variation in the peak position of $\chi_{\psibarpsi}$ as
$m$ is changed in steps of 0.01 around $m_0=0.2$.

\begin{figure}[th]
\centerline{\epsfxsize=3.2in\epsfbox{
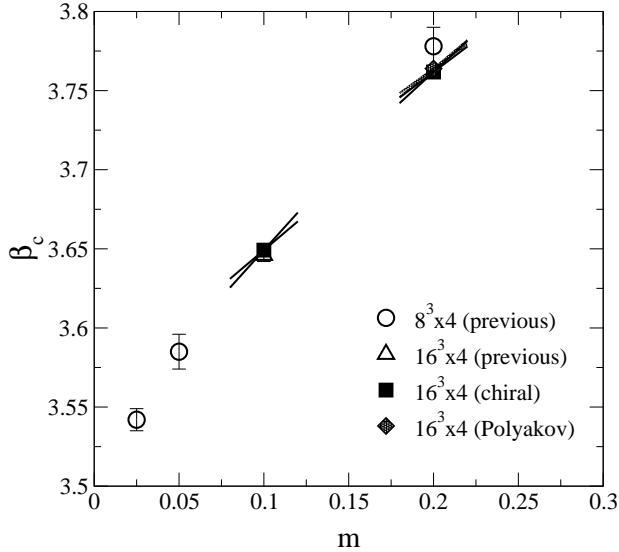
}}
\caption{The transition ``temperature'', $\beta_c$, as a function
of $m$ in comparison with previous work\protect\cite{Kar00}
\label{rew_mall}}
\end{figure}

Once these peak positions have been determined (for both
$\chi_{\psibarpsi}$ and $\chi_{L}$), $\beta_c$ can be plotted as a
function of $m$ and a comparison can be made with other
determinations. This is done in Figure \ref{rew_mall} where the
$\beta_c$ values from earlier work\cite{Kar00} are also shown.  The
line segments around our data points at $m=0.1$ and 0.2 indicate the
gradient of $\beta_c$ as a function of $m$ using our Taylor-expanded
reweighting technique. Shown in Figure \ref{rew_mall} are results from
both the chiral condensate and Polyakov loop. These are both in
perfect agreement (as expected). Furthermore, our method agrees with
previously published data confirming the validity of the approach.

%}}}

%{{{ Results for $\mu$ reweighting

\section{Results for $\mu$ reweighting}

\begin{figure}[th]
\centerline{\epsfxsize=3.2in\epsfbox{
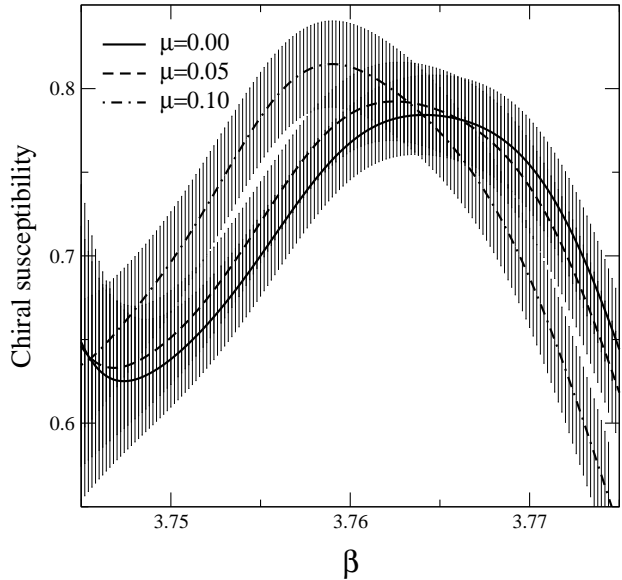
}}
\caption{The $\mu$ dependency of $\chi_{\psibarpsi}$
as a function of $\beta$
in the neighbourhood of $m=0.2$ for the chiral susceptibility.
\label{rew_mu}}
\end{figure}

We now turn to reweighting in chemical potential, $\mu$.  As in
\cite{fodorkatz}, rather than applying the method to arbitrary
parameter values, we trace out the transition point $\beta_c(\mu)$.
Using Eq.(\ref{eq:specific}), $<\psibarpsi>$ and the Polyakov Loop,
$L$, are calculated as a function of $\mu$ together with their
susceptibilities, $\chi_{\psibarpsi,L}$ (see Eq.(\ref{eq:sus})).  In
Figure \ref{rew_mu}, we plot $\chi_{\psibarpsi}$ against $\beta$ for
various $\mu$. Note that the peak position moves as $\mu$ changes.
The determination of the transition point $\beta_c$ from both the
chiral condensate and the Polyakov loop (not shown here) are found to
be in agreement.

Because we have calculated all quantities to $O(\mu^2)$ we can
extract $\beta_c$ and fit it to a quadratic in $\mu$.
(In fact, it can be shown\cite{us} that the first derivative
${\rm d}\beta_c/{\rm d}\mu =0$.)
We find ${\rm d}^2\beta_c/{\rm d}\mu^2 = 1.1 \pm 50\%$ for both
quark masses $m=0.1$ and 0.2.

We now use
\begin{eqnarray}
\frac{{\rm d}^2 T_c}{{\rm d} \mu^2} = -\frac{1}{N_t^2 T_c}
\left. \frac{{\rm d}^2 \beta_c}{{\rm d} \mu^2} \right/
\left( a \frac{{\rm d} \beta}{{\rm d} a} \right),
\end{eqnarray}
to convert ${\rm d}^2\beta_c/{\rm d}\mu^2$ into physical units,
with the beta-function ${\rm d} \beta/{\rm d} a$ coming from string
tension data\cite{Kar00}.
We find $T_c \frac{{\rm d}^2 T_c}{{\rm d} \mu^2} \approx -0.14$
at $ma=0.1$.

\begin{figure}[th]
\centerline{\epsfxsize=3.2in\epsfbox{
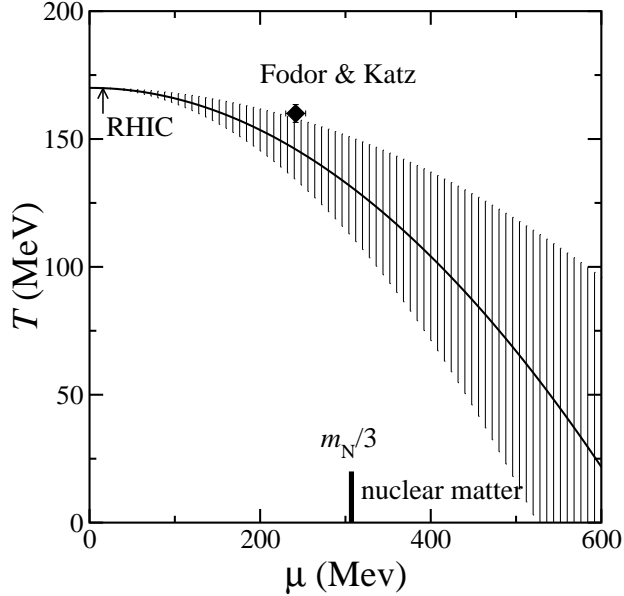
}}
\caption{The transition temperature, $T_c$, as a function
of $\mu$. The diamond symbol is the endpoint of the first order
transition obtained by Fodor and Katz\protect\cite{fodorkatz}
\label{rew_muall}}
\end{figure}

Figure \ref{rew_muall} shows the phase transition curve $T_c(\mu)$
obtained from this method. On this graph we have plotted the
Fodor-Katz point\cite{fodorkatz} which is within our errors,
confirming our method. Also shown is the $\mu$ value corresponding
to RHIC. It is interesting to extrapolate the curve $T_c(\mu)$
to the horizontal axis (as shown). It is known that the transition
(at $T\approx 0$) between ordinary hadronic matter and quark matter
occurs at around $\mu \approx 400$ MeV. This is at a smaller value
of $\mu$ than the horizontal intercept of our data indicating
(not surprisingly) the presence of higher order terms in the
Taylor expansion and/or a breakdown in our method at these large
values of $\mu$.

This motivates the question: for what range in $\mu$ do we expect our
method to be accurate (and converge to the correct answer)?  We have
studied this issue by calculating the complex phase, $\theta$, of the
fermionic determinant (which enters in the reweighting factor in
Eq.(\ref{eq:opexp})), i.e.
\begin{equation}
\det M = | \det M | {\rm e}^{i\theta}
\label{eq:theta}
\end{equation}
The reweighting method will fail when the fluctuations (standard
deviation) in $\theta$ are larger than $O(\pi/2)$.  Taylor expanding
Eq.(\ref{eq:theta}) and noting that only odd derivatives contribute to
the complex phase $\theta$, we find that the standard deviation
$\Delta\theta \sim O(\pi/2)$ at around $\mu/T_c \sim 0.5$. Since this is
around five times the $\mu$ value of RHIC, we can confirm that our
method is applicable for RHIC physics.

An interesting dynamical quark effect can be uncovered when studying
the Polyakov Loop susceptibility, $\chi_L$. For $\mu<0$, anti-quarks
are dynamically generated which screen colour charge. This leads to a
reduction in the free energy of a single quark, and a corresponding
reduction in the strength of the singularity at the transition. We
observe this effect by noting that the peak height of $\chi_L$ is
smaller for $\mu<0$ compared with $\mu>0$\cite{us}.

%}}}

%{{{ Pressure and energy density

\section{Pressure and energy density}
\label{Pressure}

Of great interest for heavy ion collision experiments is the study
of the pressure, $p$, and energy density $\epsilon$ and their $\mu$
dependence.  We can obtain estimates of $p$ by employing the integral
method\cite{integral}:
\begin{equation}
p = \frac{T}{V} \ln Z.
\end{equation}
The first derivative of $p$ w.r.t. $\mu$ is related to
the quark number density,
\begin{equation}
n_q = \frac{T}{V} \frac{ \partial \ln Z }{ \partial \mu },
\end{equation}
and the second derivative to the singlet quark number susceptibility,
$\chi_S$. Both $n_q$ and $\chi_S$ can be calculated in terms of the
quark matrix, $M$.

Using the above to estimate $p$ at the RHIC point we find that $p$
increases by around 1\% from its $\mu=0$ value.
%%
%\begin{equation}
%p({\rm RHIC}) \approx p(\mu=0) + 1 \%.
%\label{eq:p}
%\end{equation}
%%

The energy density, $\epsilon$, can be obtained from
\begin{equation}
\label{eq:e-3p}
\frac{\epsilon - 3p}{T^4}
\approx -\frac{1}{VT^3} \frac{\partial \ln {Z}}{\partial \beta}
\left( \frac{1}{a} \frac{\partial a}{\partial \beta} \right)^{-1}.
\end{equation}
The derivatives of Eq.(\ref{eq:e-3p}) can be expressed in terms of
$n_q$ and $\chi_S$.  Combining this with the above calculation of $p$,
we obtain a value for $\epsilon$ alone. We find that, at the RHIC
point, there is again only a 1\% deviation from $\epsilon(\mu=0)$.

Finally we study the variation of $p$ and $\epsilon$ along the transition
line $T_c(\mu)$. (The above calculations were performed at fixed $T$.)
Our aim is to determine whether $p$ and $\epsilon$ are constant along
$T_c(\mu)$.
The constant $p$ line is defined as
\begin{equation}
\Delta p = \frac{\partial p}{\partial T} \Delta T
+ \frac{\partial p}{\partial (\mu^2)} \Delta (\mu^2)
= 0,
\label{eq:pcons}
\end{equation}
with a similar expression for the constant $\epsilon$ line.
Using the above and the value determined earlier for the
rate of change of $T_c$ with $\mu$, we find that the
value of both $p$ and $\epsilon$ along the transition
line $T_c(\mu)$ is consistent with zero with our current
precision\cite{us}.

%}}}

%{{{ Conclusions

\section{Conclusions}

This work (which is published in full elsewhere\cite{us}) has outlined
a new method of determining thermal properties of QCD at non-zero
chemical potential from the lattice. This approach is based on Taylor
expanding the Ferrenberg-Swendsen reweighting scheme.

Using this method, the susceptibilities in the chiral condensate and
Polyakov loop were determined and their peak positions used to define
the transition point, $T_c$. As a warmup exercise, the reweighting
technique was used to determine the transition point as a function of
the quark mass, $m$ confirming earlier work. The method was then applied
to obtaining $T_c$ as a function of chemical potential, $\mu$, confirming
the work of Fodor and Katz\cite{fodorkatz}.
Very recent work of de Forcrand and Philipsen, who studied the
transition temperature for imaginary $\mu$ and then analytically
continued these results to real-valued $\mu$ also confirm
our results\cite{deFP}.

The region of applicability of our method was studied by calculating
the fluctuations in the phase of the reweighting factor. This region
was found to be substantial and easily covers the physically
interesting values of $\mu$ appropriate for RHIC physics.

We also extracted information about the pressure, $p$, and energy
density, $\epsilon$, as a function of chemical potential. We found
that the variation in these quantities from their values at $\mu=0$ is
tiny.  This leads us to conclude that RHIC physics well approximated by
$\mu=0$ physics. Furthermore we find that $p$ and $\epsilon$ are
approximately constant along the transition line $T_c(\mu)$.

The success of this work motivates the use of lighter, more
physical quark masses, and the study of (2+1) dynamical flavours
to correctly model real world physics.

%}}}

%{{{ acknowledgements

\section*{Acknowledgments}

The authors would like to acknowledge the Particle Physics and Astronomy
Research Council for the award of the grant PPA/G/S/1999/00026 and
the European Union for the grant ERBFMRX-CT97-0122.
CRA would like to thank the Department of Mathematics, University of
Queensland, Australia for their kind hospitality while part of this work was
performed.

%}}}

%{{{ bibliography

%}}}

\end{document}